\documentclass[twocolumn,aps,longbibliography,floatfix,showpacs,superscriptaddress]{revtex4}
\usepackage{amsmath}
\usepackage{cases}
\usepackage[dvipdfmx]{graphicx}
\usepackage{braket}
\usepackage{bm}

\def\d{{\partial}}

\def\e{{\epsilon}}
\def\k{{ {\bm k} }}

\def\q{{ {\bm q} }}
\def\Q{{ {\bm Q} }}
\def\0{{ {\bm 0} }}

\def\w{{\omega}}
\def\a{{\alpha}}

\allowdisplaybreaks[4]

\begin{document}
\title{
Self-energy driven resonance-like inelastic neutron spectrum 
in $s_{++}$-wave state in Fe-based superconductors
}
\author{Lisa \textsc{Takeuchi}, Youichi \textsc{Yamakawa}, 
 and Hiroshi \textsc{Kontani}}
\date{\today}

\begin{abstract}
To elucidate the pairing states in Fe-based superconductors,
we perform careful calculation of 
the dynamical spin susceptibility $\chi^S(\q,\w)$ 
at very low temperatures ($T\gtrsim1$meV).
The feedback effect on both the self-energy and $\chi^S(\q,\w)$ 
from the superconducting gap are self-consistently analyzed
based on the fluctuation-exchange (FLEX) approximation.
In the $s_{\pm}$-wave state, which has sign-reversal in the gap function,
$\chi^S(\q,\w)$ at the nesting momentum $\q=\Q$
shows a resonance peak even when
the system is away from the magnetic quantum-critical-point (QCP).
In the $s_{++}$-wave state that has no sign-reversal,
$\chi^S(\q,\w)$ shows a large hump structure when the system is 
close to the magnetic QCP.
This result confirms the validity of 
self-energy driven resonance-like peak in $s_{++}$-wave state
proposed in our previous semi-microscopic study:
The enhancement in $\chi^S(\q,\w)$ 
due to self-energy effect exceeds the 
suppression due to coherence factor effect near magnetic QCP.
We stress that the hump structure in the $s_{++}$-wave state
given by the FLEX method
smoothly changes to resonance-like sharp peak structure
as the system approaches magnetic QCP,
which was not reported in our previous studies.
The obtained $\w$- and $T$-dependences of $\chi^S(\q,\w)$ 
in the $s_{++}$-wave state resemble to 
the resonance-like feature in inelastic neutron scattering spectra 
recently observed in Na(Fe,Co)As and FeSe.

\end{abstract}

\address{
Department of Physics, Nagoya University,
Furo-cho, Nagoya 464-8602, Japan. 
}


\sloppy

\maketitle
\section{Introduction}

In Fe-based superconductors,
the pairing mechanism and the gap structure with 
$s$-wave symmetry have been central open problems.
When inter-pocket repulsive interaction 
due to the spin fluctuations is strong, 
the fully-gapped sign-reversing $s$-wave state ($s_\pm$-wave state)
is expected to appear \cite{Kuroki,Mazin}.
On the other hand, 
when orbital-fluctuation-driven inter-pocket attractive 
interaction is strong, the fully-gapped $s$-wave state 
without sign-reversal ($s_{++}$-wave state) will emerge 
\cite{Kontani-Onari,onari_anion,yama_holepocket}.
In many optimally-doped Fe-based superconductors, 
nematic orbital fluctuations and spin fluctuations 
develop cooperatively, as reported in Refs. 
\cite{yoshizawa,gallais}.
Theoretically, strong orbital fluctuations are driven by 
moderate spin fluctuations, thanks to the
orbital-spin mode-coupling due to the higher-order
many-body effects, especially the Aslamazov-Larkin type
vertex correction (AL-VC)
\cite{onari_c2symm,yamakawa-fese,onari_scvc,kontani_c66,chubukov2018}.

To detect the presence or absence of the 
sign-reversal in the gap,
phase-sensitive experiments are very significant.
Nonmagnetic impurity effect provides us useful
phase-sensitive information.
In typical $d$-wave superconductors,
like cuprate superconductors and CeCoIn$_5$, $T_{\rm c}$ is quickly 
suppressed by impurities, following the prediction of the 
Abrikosov-Gor'kov theory
 \cite{Yamashita-imp-exp}.
In many Fe-based superconductors,
the superconductivity survives even when the 
residual resistivity due to the randomness is very high,
comparable to high-$T_{\rm c}$ $s$-wave superconductors
MgB$_2$ and YNi$_2$B$_2$C
 \cite{Li-impurity-exp,Yamashita-imp-exp}.
Since the $s_{\pm}$-wave state is as fragile 
as the $d$-wave state theoretically
\cite{onari-impurity,yama_impurity},
these experiments support the (impurity-induced) 
$s_{++}$-wave state in optimally-doped pnictides
\cite{Kontani-Onari,Efremov-imp,Korshunov-imp}.

Another promising phase-sensitive experiment 
is the inelastic neutron scattering study.
Large resonance peak in the dynamical spin susceptibility
appears in $d$-wave superconductors, 
such as cuprates \cite{iikubo-sato,ito-sato,keimer-highTc}
and CeCoIn$_5$ \cite{stock-CeCoIn5},
reflecting the sign-reversal of the $d$-wave gap
\cite{Monthoux-Scalapino,pines,chubukov-resonance,takimoto-moriya}.
In the $s_\pm$-wave state,
$\chi^S(\q,\w)$ is expected to show the resonance peak at 
$\w=\w_{\rm res}<2\Delta$ since 
the coherence factor enlarges the spin fluctuation for 
$s_\pm$-wave state
\cite{maier-scalapino,eremin,Das-res,Kuroki-res,Korshunov-res,Korshunov-res2},
while it suppresses the spin fluctuation for $s_{++}$-wave state.
($\Delta$ is the amplitude of the gap function.)
Experimentally, clear broad peak structures in $\chi^S(\q,\w)$
were observed for $T\ll T_{\rm c}$ in 
FeSe \cite{neutron-FeSe}, 
BaFe$_{2-x}$Co$_{x}$As$_2$ \cite{keimer,tate},
Ca-Fe-Pt-As \cite{Sato-neutron}, 
Na(Fe,Co)As \cite{neutron-NaFeAs},
and (Ba,K)Fe$_2$As$_2$ \cite{lee-resonance}.
However, $\chi^S(\q,\w)$ is drastically modified by not only 
the coherence factor, but also by the self-energy effect.
In fact, experimentally observed hump structures
can be explained based on the $s_{++}$-wave state
\cite{onari-resonance,onari-resonance2},
if one considers the energy-dependence of the 
normal self-energy $\Sigma(\k,\w)$.
This effect is totally dropped in the random-phase-approximation (RPA).
Previous theoretical studies
\cite{onari-resonance,onari-resonance2} claim that 
the peak energy $\w_{\rm res}$ of the hump structure
in the $s_{++}$-wave state
satisfies the relation $\w_{\rm res}\gtrsim 2\Delta$.

To distinguish between the resonance peak 
and the hump structure experimentally,
it is important to verify the resonance condition $\w_{\rm res}< 2\Delta$.
However, it is very difficult to 
obtain the accurate gap amplitude $\Delta$ experimentally.
In addition, from the theoretical viewpoint, 
we cannot rule out the relation $\w_{\rm res}< 2\Delta$
in the $s_{++}$-wave state if the system is very close to 
the magnetic quantum-critical-point (QCP),
as we will discuss in this paper.

The main player in realizing the hump structure 
of the $s_{++}$-wave state is the $\w$-dependence 
of the inelastic quasiparticle damping 
$\gamma_\k^*(\w) \equiv -{\rm Im} \Sigma^R(\k,\w) / Z(\k,\w)$,
where $Z(\k,\w)$ is the mass-enhancement factor.
Above $T_{\rm c}$, $\chi^S(\q,\w)$ is strongly suppressed 
by large $\gamma_\k^*(\w)$.
Since $\gamma_\k^*(\w)\approx0$ for $\w<3\Delta$ for $T\ll T_{\rm c}$,
$\chi^S(\q,\w)$ takes large hump structures
at $ \w \gtrsim 2\Delta$.
Nonetheless of the significance of the $\gamma_\k^*(\w)$,
the authors in Refs. \cite{onari-resonance,onari-resonance2}
assumed a very simple functional form of $\gamma_\k^*(\w)$,
just as a phenomenological function.
In addition, the renormalization effect due to the real part of
 $\Sigma(\k,\w)$ was dropped.
In order to verify the hump-structure mechanism 
in the $s_{++}$-wave state without ambiguity,
the self-consistent calculation between 
$\chi^S(\q,\w)$ and $\Sigma(\k,\w)$
should be performed at sufficiently low temperatures.


In this paper,
we study the dynamical spin susceptibility $\chi^S(\q,\w)$ 
in order to elucidate the pairing states in Fe-based superconductors.
We self-consistently calculate $\chi^{S}(\q,\w)$ and the normal self-energy 
$\Sigma(\k,\w)$ using the fluctuation-exchange (FLEX) approximation
\cite{takimoto-moriya,bickers,pao,monthoux,dahm,langer,grabowski,deisz,putz,esirgen,koikegami,kita}.
We develop the multi-step FLEX procedure to perform
precise numerical studies at very low temperatures ($T\gtrsim1$meV), 
and analyze the feedback effect on $\Sigma(\k,\w)$
and $\chi^S(\q,\w)$ from the superconducting gap carefully.
In the $s_{\pm}$-wave state,
$\chi^S(\q,\w)$ shows the resonance peak even when
the system is away from the magnetic QCP.
In the $s_{++}$-wave state, $\chi^S(\q,\w)$ shows 
large hump structures near the magnetic QCP,
since the enhancement in $\chi^S(\q,\w)$ 
due to self-energy effect exceeds the 
suppression due to coherence factor effect.
This result confirms the validity of self-energy driven resonance-like peak 
in $s_{++}$-wave state, which was proposed in our 
previous semi-microscopic study 
\cite{onari-resonance,onari-resonance2}.
We stress that the hump structure in $s_{++}$-wave state 
smoothly changes to resonance-like sharp peak
as the system approaches to magnetic QCP,
which was not reported previously
\cite{onari-resonance,onari-resonance2}.
The obtained $\w$- and $T$-dependence of $\chi^S(\q,\w)$ 
in the $s_{++}$-wave states near the magnetic QCP 
resemble to the inelastic neutron scattering spectra 
in Na(Fe,Co)As \cite{neutron-NaFeAs} and 
FeSe \cite{neutron-FeSe}.

Mathematically, resonance peak appears
in case that the dynamical spin Stoner factor $\a_S(\w)$ 
reaches unity for $\w<2\Delta$.
We show that $\a_S(\w_{\rm res})\approx1$ is realized
even in the $s_{++}$-wave state near magnetic QCP,
if $(\w,T)$-dependence of the self-energy 
is taken into account correctly.


\section{Model}

\subsection{Hubbard model}
\begin{figure}[htb]
  \centering
  \includegraphics[width=0.7\linewidth]{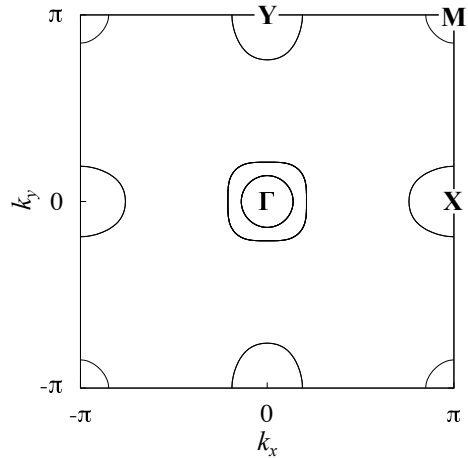}
  \caption{Fermi surface of 1111-type iron-based superconductors.
	}
  \label{fermisurface}
\end{figure}

The Hamiltonian used is the 2-dimensional 5-orbital Hubbard model \cite{onari-impurity}
\begin{eqnarray}
  H =  \sum_{ij} \sum_{lm} \sum_{\sigma} t_{ij}^{lm}c_{il\sigma}^{\dagger}c_{jm\sigma} + H_{{\rm Coulomb}}, 
\end{eqnarray}
where $i,j$ are the Fe sites, $l,m$ represent the $d$-orbitals, and $\sigma$ is the spin index. 
The interaction potentials included are intra-orbital Coulomb potential $U$, inter-orbital Coulomb potential $U'$, Hund's coupling $J$, and pair hopping $J'$.
The hopping parameters used are those of 1111-type iron-based superconductors, and the Fermi surface obtained is in Fig. \ref{fermisurface}.

Using this Hamiltonian, we solve the 10$\times$10 Nambu Green's function for superconducting state in orbital representation,
\begin{equation}
  \widehat{\mathcal{G}}(k) =
  \begin{pmatrix}
    \widehat{G}_(k)  &  \widehat{F}(k)\\
    \widehat{F}^{\dagger}(k)  &  -\widehat{G}^{t}(-k)
  \end{pmatrix} ,
\end{equation}
where $k=(\bm{k},i\epsilon_n)$.
This can be calculated from finding the inverse of the following matrix,
\begin{eqnarray}
  \widehat{\mathcal{G}}^{-1}(k) =
  i\epsilon_n \widehat{1} -
  \begin{pmatrix}
    \widehat{H}(\k) + \widehat{\Sigma}(k) & \widehat{\Delta}(\bm{k})\\
    \widehat{\Delta}^\dagger(\bm{k}) & - \widehat{H}^{t}(-\k) - \widehat{\Sigma}^{t}(-k)
  \end{pmatrix},
\label{eqn:Green}
\end{eqnarray}
where 
$\widehat{1}$ is the identity matrix, 
$\widehat{\Delta}$ is the superconducting gap without renormalization, 
and $\epsilon_n = \pi T (2n+1)$ is the Matsubara frequency for fermions. 
$\widehat{\Sigma}$ is the normal self-energy, which represents the 
mass-enhancement and quasiparticle damping.
In this study, we introduce $\widehat{\Delta}$ as a parameter.

\subsection{Gap functions}

To calculate the spin susceptibility in superconducting state, 
we introduce the ``unrenormalized gap functions'' 
in Eq. (\ref{eqn:Green}).
In each band,
we introduce the following $s_{++}$- and $s_{\pm}$-wave gap functions in band representation: 
\begin{eqnarray}
   s_{++}  &:& \Delta(\bm{k}) = \Delta_0 \\
   s_{\pm} &:& \Delta(\bm{k}) = \Delta_0[\cos(k_x)+\cos(k_y)].
 \label{eqn:D0}
\end{eqnarray}
To calculate the temperature dependence of spin susceptibility,
we introduce a superconducting gap with a BCS-like temperature dependence,
$\Delta_0(T) = \Delta_0(0)\tanh{\left(1.74\sqrt{(T_{\rm c}/T)-1}\right)}$.

Hereafter, in the numerical study using the FLEX approximation, 
we set $T_{\rm c}=8$meV and $\Delta_0(T=0) = 50$meV
unless otherwise noted.
The physical gap function is given as 
$\Delta^*(\k) \approx \Delta(\k)/Z(\k)$, where
$Z(\bm{k}) =1-\frac{\d}{\d\w}\Sigma(\k,\w)|_{\w=0}$ 
is the mass-enhancement factor given by the normal self-energy.

In optimally-doped Fe-based superconductors, 
the energy-scale of spin and/or orbital fluctuations,
which gives the pairing glue, is small.
For this reason, $\Delta(\bm{k})$ should be large only near the Fermi level.
To express this fact
\cite{Maier},
we introduce the following high-energy cutoff for the gap function:
\begin{equation}
  \Delta_{e}(\bm{k}) = \Delta(\bm{k}) \cdot \frac{\epsilon_{\rm cut}^2}{[\epsilon(\bm{k})-\mu]^2+\epsilon_{\rm cut}^2},
\end{equation}
with $\epsilon_{\rm cut} = 4\Delta_0$. 
The gap function in orbital representation can be expressed as
\begin{eqnarray}
  \left[\widehat{\Delta}(\bm{k})\right]_{lm} = \sum_{b}U_{lb}(\bm{k})U^{\dagger}_{mb}(\bm{k})\Delta_e(\bm{k}),
\end{eqnarray}
where $U_{lb} (\bm{k})$ is the unitary matrix element 
and $b$ represents the conduction band number. 

\subsection{FLEX approximation}
The FLEX approximation is a method to calculate self-energy 
and susceptibilities in a self-consistent manner.
Feedback effect from spin fluctuation is included by using FLEX approximation,
allowing a microscopic calculation of the system.

The bare susceptibilities in Matsubara frequency representation are written as
\begin{eqnarray}
  \chi^0_{ll'mm'} (q)
  &=& -\frac{T}{N}\sum_{k} G_{lm}(k+q) G_{m'l'}(k),\\
  \phi^0_{ll'mm'} (q)
  &=& -\frac{T}{N}\sum_{k} F_{lm'}(k+q) F^{\dagger}_{l'm}(k),
\label{phi0}
\end{eqnarray}
where $q=({\bm q},i\omega_l)$, $\omega_l=2\pi l T$, and $N$ is the number of $\k$-mesh. The spin and charge susceptibilities are
\begin{eqnarray}
  \chi^S_{ll'mm'}(q)
  &=& \left[\frac{\widehat{\chi}^{0}(q)+\widehat{\phi}^{0}(q)}
  {1-\widehat{\Gamma}^S[\widehat{\chi}^{0}(q)+\widehat{\phi}^{0}(q)]}\right]_{ll'mm'},\\
  \chi^C_{ll'mm'} (q)
  &=& \left[\frac{\widehat{\chi}^{0}(q)-\widehat{\phi}^{0}(q)}
  {1-\widehat{\Gamma}^C[\widehat{\chi}^{0}(q)-\widehat{\phi}^{0}(q)]}\right]_{ll'mm'},
\end{eqnarray}
where $\widehat{\Gamma}^S$ and $\widehat{\Gamma}^C$ are the spin and charge interaction matrices, respectively. \cite{yamakawa-fese}

The Feynman diagrams considered in the calculation of self-energy are the bubble terms and ladder terms.
\begin{widetext}
Normal self-energy in the superconducting state is calculated by the expression
\begin{equation}
  \Sigma_{lm}(k)=\frac{T}{N}\sum_{q}\sum_{l'm'}G_{l'm'}(k-q)V_{ll'mm'}(q) .
\end{equation}
The interaction part $V_{ll'mm'}(q)$ is
\begin{eqnarray} 
  V_{ll'mm'}(q) &=&  \frac{3}{2}V^S_{ll'mm'}(q)+\frac{1}{2}V^C_{ll'mm'}(q) \nonumber\\
  -\sum_{l_1l_2l_3l_4} &\Bigl[&
    +\frac{1}{4}\Gamma^{\uparrow\uparrow}_{ll'l_1l_2} \chi^0_{l_1l_2l_3l_4}(q) \Gamma^{\uparrow\uparrow}_{l_3l_4mm'}
    +\frac{1}{2}\Gamma^{\uparrow\downarrow}_{ll'l_1l_2} \chi^0_{l_1l_2l_3l_4}(q) \Gamma^{\uparrow\downarrow}_{l_3l_4mm'} \nonumber \\
  &&-\frac{1}{2}\Gamma^{\uparrow\uparrow}_{ll'l_1l_2} \phi^0_{l_1l_2l_3l_4}(q) \Gamma^{\uparrow\downarrow}_{l_3l_4mm'}
  -\frac{1}{2}\Gamma^{\uparrow\downarrow}_{ll'l_1l_2} \phi^0_{l_1l_2l_3l_4}(q) \Gamma^{\uparrow\uparrow}_{l_3l_4mm'} \Bigr]. 
\end{eqnarray}
Here we defined
$\widehat{V}^{S(C)} = \widehat{\Gamma}^{S(C)} \widehat{\chi}^{S(C)} \widehat{\Gamma}^{S(C)}$, 
$\widehat{\Gamma}^{\uparrow\uparrow} = (\widehat{\Gamma}^C + \widehat{\Gamma}^S)/2$, and 
$\widehat{\Gamma}^{\uparrow\downarrow} = (\widehat{\Gamma}^C - \widehat{\Gamma}^S)/2$.
For the numerical study of the spin susceptibility,
we derive the retarded (advanced) self-energy 
$\widehat{\Sigma}^{R(A)}(\k,\e)$
from $\widehat{\Sigma}(\k,i\e_n)$ given by the FLEX
by performing the numerical analytic continuation.

\end{widetext}

\subsection{Spin susceptibility}

The normal bare susceptibility $\chi^{0,R}$
and the anomalous bare susceptibility $\phi^{0,R}$ in the real energy representation
can be expressed by the following equations.\cite{takimoto-moriya,onari-resonance2}
\begin{eqnarray}
  &\chi^{0,R}_{ll'mm'}&(\q,\omega)
  = \frac{-1}{4\pi iN}\sum_{\k}\nonumber\\
    &&\Bigl[\int_{-\infty}^{\infty} dz \tanh\left(\frac{z}{2T}\right)G_{lm}^{R}(\k^+,z^+)\rho^{G}_{m'l'}(\k,z)\nonumber\\
    && +  \int_{-\infty}^{\infty} dz \tanh\left(\frac{z}{2T}\right)\rho_{lm}^{G}(\k^+,z^+)G^{A}_{m'l'}(\k,z) \Bigr], \nonumber\\
  \label{chi_r}
\end{eqnarray}
\begin{eqnarray}
  &\phi^{0,R}_{ll'mm'}&(\q,\omega)
  = \frac{-1}{4\pi iN}\sum_{\k} \nonumber\\
    &&\Bigl[\int_{-\infty}^{\infty} dz \tanh\left(\frac{z}{2T}\right)F_{lm'}^R(\k^+,z^+)\rho^{F^{\dagger}}_{ml'}(\k,z)\nonumber\\
    && +  \int_{-\infty}^{\infty} dz \tanh\left(\frac{z}{2T}\right)\rho_{lm'}^{F}(\k^+,z^+)F^{\dagger A}_{ml'}(\k,z) \Bigr]. \nonumber\\
  \label{phi_r}
\end{eqnarray}
Here, $\rho^G_{ll'} = (G^A_{ll'} - G^R_{ll'})/2\pi i$
and $\rho^{F^{(\dagger)}}_{ll'} = (F^{(\dagger)A}_{ll'} - F^{(\dagger)R}_{ll'})/2\pi i$.
$G^A$, $F^A$ are the advanced Green's functions,
and $G^R$, $F^R$ are the retarded Green's functions.
We defined $\k^+ = \k + \q$ and $z^+ = z+\omega$ for simplicity. 

The spin susceptibility $\chi^{S,R}$ can be expressed by the equation
\begin{eqnarray}
  \chi^{S,R}_{ll'mm'}(\q,\omega)
  = \left[\frac{\widehat{\chi}^{0,R}(\q,\omega)+\widehat{\phi}^{0,R}(\q,\omega)}
  {1-\widehat{\Gamma}^S[\widehat{\chi}^{0,R}(\q,\omega)+\widehat{\phi}^{0,R}(\q,\omega)]}\right]_{ll'mm'}.
\end{eqnarray}
Here, we introduce the Stoner factor $\a_S$ defined as
the maximum eigenvalue of 
\begin{equation}
  \widehat{\Gamma}^S[\widehat{\chi}^{0,R}(\q,0)+\widehat{\phi}^{0,R}(\q,0)].
\end{equation}
It is proportional to the strength of spin fluctuation; $\chi^S$ diverges when $\alpha_S$ is 1.

The results of neutron-scattering experiments corresponds to the imaginary part of spin susceptibility,
\begin{equation}
  {\rm Im}\chi^S(\Q,\omega) = {\rm Im}\left[\sum_{lm}\chi^{S,R}_{llmm}(\Q,\omega)\right] .
\end{equation}

\section{Results}

In order to calculate at low temperatures,
multi-step method is used in this research. 
We present the explanation for this method in Appendix A.
Results for FLEX approximation and RPA are calculated
with a $\k$-mesh of $128^2$ and Matsubara frequency of $2^{16}$.
Bare susceptibilities in energy representation (Eqs. \ref{chi_r} and \ref{phi_r}) are calculated
with $\k$-mesh of $256^2$ and energy range divided by $2^{12}$ ($\delta z \sim 1$meV).

In this section, 
we perform self-consistent numerical study
based on the FLEX approximation.
Except for Fig. \ref{comparison},
we calculate $T$-dependences of physical quantities,
for a fixed Coulomb interaction
which satisfy the condition $\a_S=0.90\sim0.97$ 
at $T=T_{\rm c}(=8\ {\rm meV})$.

\subsection{Feedback Effect}

\begin{figure}[htb]
  \centering
  \includegraphics[width=0.8\linewidth]{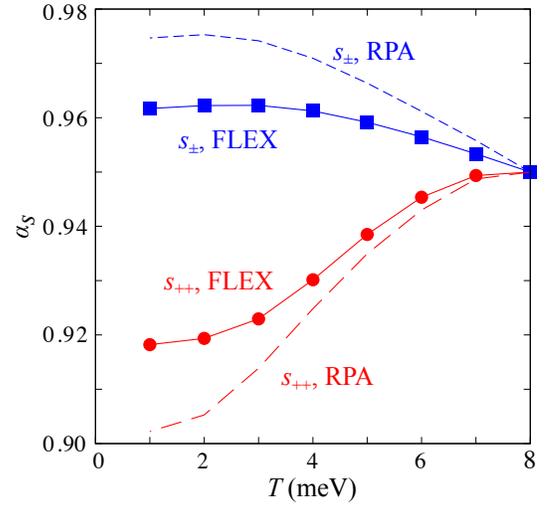}
  \caption{$T$-dependence of Stoner factor at nesting vector $\bm{Q}$
given by the RPA ($\Delta_0=10$ meV) 
and by the FLEX ($\Delta_0=50$ meV).
In each case, $\alpha_S=0.95$ at $T_{\rm c}$.
We set $U_{\rm FLEX}=2.11$ eV and $U_{\rm RPA}=1.13$ eV.
Note that $\Delta^*\approx 13 \ (9.6)$ meV 
for $s_{++}$ ($s_\pm$)-wave state as shown in Fig. \ref{renorm}.
}
  \label{feedback}
\end{figure}

Figure \ref{feedback} shows the $T$-dependence of Stoner factor $\a_S$
in the superconducting state.
The Stoner factor behaves differently for $s_{++}$ 
and $s_{\pm}$ states, in both RPA and FLEX approximation.
One of the reasons is that 
$\phi^0(\q)$ in the irreducible susceptibility is
proportional to $-\Delta(\k)\Delta(\k+\q)$,
which is negative (positive) for 
$s_{++}$-wave ($s_{\pm}$-wave) states
at nesting vector $\Q=(\pi,0)$.
By reflecting the difference in sign of this factor,
which corresponds to the difference in the coherence factor
in the BCS theory,
$\a_S$ in the $s_{++}$-wave state is smaller than 
$\a_S$ in the $s_\pm$-wave state.
To summarize, 
$\a_S$ slightly decreases in the $s_{++}$-wave state,
whereas it increases in the $s_\pm$-wave state.

The $T$-dependence of $\a_S$ obtained by the RPA for $\Delta_0=10$meV 
is similar to that obtained by the FLEX for $\Delta_0=50$meV,
for both $s_{++}$ and $s_\pm$ states.
This result is reasonable because 
the renormalized gap in the FLEX 
averaged over the Fermi surfaces
is $\Delta^*\approx 13 \ (9.6)$ meV 
for $s_{++}$ ($s_\pm$)-wave state at $T=1$meV 
as shown in Fig. \ref{renorm}.

\begin{figure}[htb]
  \centering
  \includegraphics[width=0.9\linewidth]{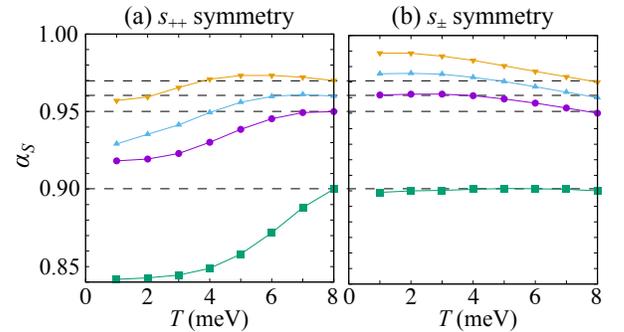}
  \caption{
$T$ dependence of $\a_S$ for the initial values
$\alpha_S=0.90\sim0.97$ at $T=T_{\rm c}$,
in the case of (a) $s_{++}$-wave state and (b) $s_\pm$-wave state.
}
  \label{feedback_comp}
\end{figure}

Figure \ref{feedback_comp} shows $\a_S$ in the superconducting state
obtained by the FLEX approximation for $\Delta_0=50$meV,
in the case of $\alpha_S(T_{\rm c})=0.90\sim0.97$.
In Fig. \ref{feedback_comp} (a) for the $s_{++}$-wave state, 
the Stoner factor monotonically decreases with 
decreasing $T$ in the case of $\alpha_S(T_{\rm c})\le 0.95$.
In contrast, the Stoner factor first increases slightly
and then decreases at low temperatures
in the case of  $\alpha_S(T_{\rm c})\ge 0.96$.
In Fig. \ref{feedback_comp} (b) for the $s_\pm$-wave state, 
the Stoner factor is almost constant when $\alpha_S(T_{\rm c})\le 0.90$.
In contrast, the Stoner factor monotonically increases 
with decreasing $T$ when $\alpha_S(T_{\rm c})\ge 0.95$.

Thus, when the system is close to the magnetic QCP in the normal state,
the spin fluctuations remain strong 
even in the $s_{++}$-wave superconducting states.
The reason is the following:
in the normal state at $T \geq T_{\rm c}$,
$\chi^S(\q,\w=0)$ is suppressed by the large inelastic scattering 
$\gamma_\k(\w)$ for $\w \sim 0$.
(Here, $\gamma_\k(\w)$ is the imaginary part of the self-energy.)
For $T\ll T_{\rm c}$,
$\gamma_\k(\w)$ is prominently reduced for $|\w| < 3\Delta^*$
(see Fig. \ref{damping}), which leads to the increment of 
$\chi^S(\q,\w=0)$.
Therefore, the self-energy gives the positive feedback
from the superconducting gap to the spin susceptibility,
for both $s_{++}$ and $s_\pm$ states.
To summarize, both the coherent factor and the self-energy effect
are important for understanding the spin fluctuations
in the superconducting state.

\subsection{Renormalized gap size $\Delta^{*}$}
\begin{figure}[htb]
  \centering
  \includegraphics[width=0.7\linewidth]{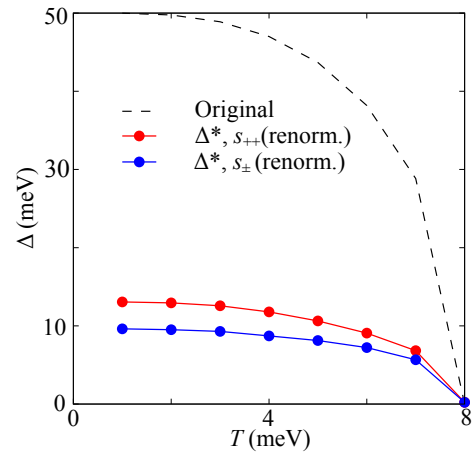}
  \caption{Renormalized superconducting gap $\Delta^*$
given by the FLEX approximation, 
for both $s_{++}$ and $s_{\pm}$-wave states.
The original unphysical gap $\Delta_0$ is also plotted.
}
  \label{renorm}
\end{figure}

By including the normal self-energy,
the original gap $\Delta_0$ in Eq. (\ref{eqn:D0})
is renormalized to be the physical gap function $\Delta^*$.
Figure \ref{renorm} shows both $\Delta_0$ and $\Delta^{\ast}$
obtained by the FLEX approximation; $\a_S(T_{\rm c})=0.95$.
The size of $\Delta^*$ is estimated numerically from the relation
$1/{\rm Re}F(\k,\w)=0$ on the Fermi surfaces.
In the $s_{++}$-wave state,
$2\Delta^{\ast}\approx13$meV, so the relation 
$2\Delta^{\ast}/T_{\rm c} \approx 3.3$ holds at $T=1$meV.
The ratio $2\Delta^{\ast}/T_{\rm c}$ increases to 4.3
if we set $T_{\rm c}=6$meV.
We remark that 
the numerical result of $\chi^S(\q,\w)$ 
is insensitive to $T_{\rm c}$ in case $T\lesssim 0.5 T_{\rm c}$.

\begin{figure}[htb]
  \centering
  \includegraphics[width=0.7\linewidth]{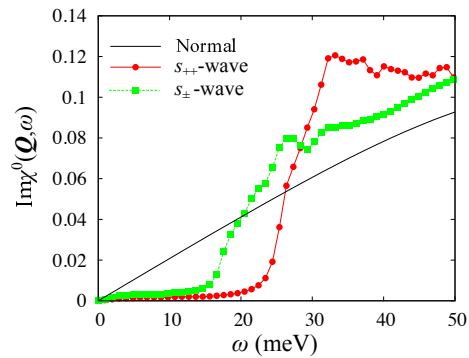}
  \caption{$\w$-dependence of ${\rm Im}\chi^0(\Q,\omega)$ at $T=3$meV
for both $s_{++}$ and $s_{\pm}$ states.
In each state, we set $\a_S(T_{\rm c})=0.95$.
The result for the normal state ($T=T_{\rm c}=8$meV) is plotted for comparison. 
$\Delta^*\approx13$meV ($9.6$meV)
for $s_{++}$-wave ($s_{\pm}$-wave).
}
  \label{imchi0}
\end{figure}

We can also derive $\Delta^*$ from 
the energy dependence of ${\rm Im}\chi^0(\Q,\omega)$.
Since ${\rm Im}\chi^0(\Q,\omega)$ is the absorption spectrum of 
particle-hole scattering, it should be zero for $|\w|<2\Delta^{\ast}$
at zero temperatures.
Figure \ref{imchi0} shows the ${\rm Im}\chi^0(\Q,\omega)$ 
obtained by the FLEX at $T=3$meV, 
for both $s_{++}$ and $s_{\pm}$-wave states.
From the result, $\Delta^{\ast}$ is estimated to be 
$10\sim15$meV for both $s_{++}$-wave and $s_{\pm}$-wave state,
consistent with the results in Fig. \ref{renorm}.

\subsection{Damping $\gamma$ due to inelastic scattering}

\begin{figure}[htb]
  \centering
  \includegraphics[width=0.9\linewidth]{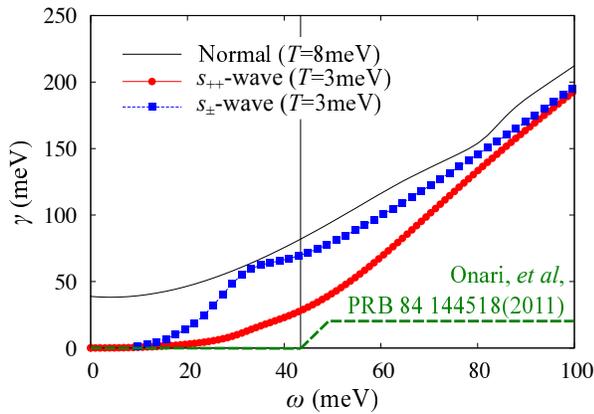}
  \caption{Energy dependence of $\gamma(\k,\w)$ 
at $\k=(0.74\pi,0)$ given by the FLEX approximation.
We set $\alpha_S=0.95$ in the normal state at $T=T_{\rm c}=8$meV.
In this case, Fermi liquid behavior 
$\gamma(\w)\approx\gamma_0+ a \w^2$ is obtained.
The low-energy ($\w\sim0$) inelastic scattering is strongly 
suppressed in both $s_{++}$ and $s_{\pm}$-wave states. 
Oversimplified phenomenological damping rate 
introduced in the previous research 
\cite{onari-resonance2}
is plotted for comparison.
Note that the renormalized quasiparticle damping is $\gamma^* = \gamma / Z$,
where $Z$ is the mass-enhancement factor ($Z \sim 5$).
}
  \label{damping}
\end{figure}

Figure \ref{damping} shows the energy dependence of 
quasiparticle damping,
$ \gamma(\k,\omega) = - \sum_{l} {\rm Im}(\Sigma^R_{ll}(\k,\omega))$,
given by the FLEX approximation.
Compared to the normal state ($T=T_{\rm c}$), the damping in the 
superconducting state ($T=3$meV) is 
drastically suppressed for the lower energy region.
In the $s_{++}$-wave state,
$\gamma(\k,\omega)$ is suppressed for $|\w|<3\Delta^*$
 \cite{onari-resonance2}.
The reason is as follows:
in the inelastic scattering process,
the initial quasiparticle with energy $E_i$ should 
create a particle-hole excitation with $E_{ph}>2\Delta^*$,
and the final quasiparticle should satisfy $E_f>\Delta^*$.
Thus, the relation $E_i=E_{ph}+E_f>3\Delta^*$ is required at $T=0$.
In the $s_\pm$-wave state,
$\gamma(\k,\w)$ is large even for $|\w|\lesssim 3\Delta^*$
since the low-energy collective resonance mode 
($\w_{\rm res}<2\Delta^*$; see Fig. \ref{resonance})
contributes to the low-energy inelastic scattering processes.

\subsection{Calculations for Neutron-scattering experiment}

Here, we explain the dynamical susceptibility $\chi^S(\q,\w)$
obtained by the FLEX approximation for various parameters,
in the cases of $s_{++}$-wave and $s_\pm$-wave states.
The resonance-like hump structure in the $\w$-dependence of 
$\chi^S(\q,\w)$ appears even in the $s_{++}$-wave state.
This result is consistent with previous RPA analysis
in Refs. \cite{onari-resonance,onari-resonance2}.

\subsubsection{$\a_S$ dependence of Im$\chi^S(\q,\w)$}
\begin{figure}[htb]
  \centering
  \includegraphics[width=0.7\linewidth]{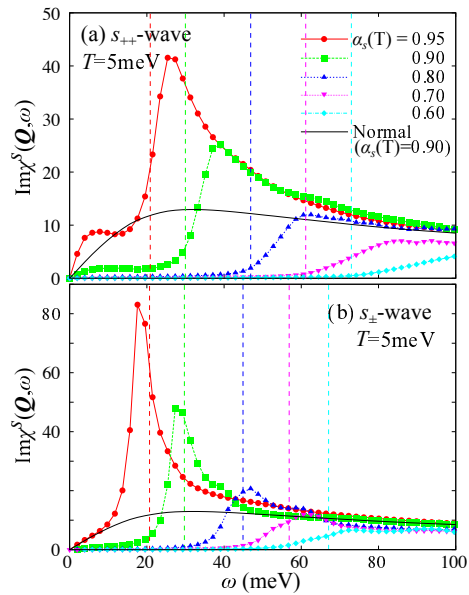}
  \caption{Energy dependence of ${\rm Im}\chi^S$ for (a)$s_{++}$ and (b)$s_{\pm}$-wave states for $\alpha_S(T)=0.95 \sim 0.6$ at $T=5$meV.
    Each calculation is in superconducting state 
    with $\Delta_0=50$meV.
The value of $\Delta^{*}$ for each $\a_S(T)$
is shown by vertical dotted line.
}
  \label{comparison}
\end{figure}

Figure \ref{comparison} shows the $\w$-dependence of 
${\rm Im}\chi^S(\Q,\w)$ for various Stoner factors,
in both (a) $s_{++}$-wave and (b) $s_\pm$-wave states.
Calculations are done with parameters $T=5$meV and $\Delta_0=50$meV.
The nesting vector is $\Q = (\pi, 0)$.
In Fig. \ref{comparison}(a) we find that a hump structure is 
obtained in the $s_{++}$-wave state, even for moderate 
spin fluctuations ($\a_S=0.90$).
The obtained hump structure is similar to the report of 
previous RPA analysis \cite{onari-resonance2}.
As the spin fluctuations become stronger ($\a_S = 0.95$), 
the hump structures become narrower 
such that it looks like a resonance peak.
The relation $\w_{\rm res} \approx 2\Delta^*$ is satisfied.
In Fig. \ref{comparison}(b), we see that unlike $s_{++}$ state, 
results for $s_{\pm}$ state have a peak-like structure at small spin fluctuation.
The condition of the resonance $\w_{\rm res}<2\Delta^*$
is apparently satisfied for $\a_S\ge0.9$.
As $\a_S$ is increased, the resonance peak becomes narrower
and $\w_{\rm res}$ decreases.

\subsubsection{Temperature dependence of Im$\chi^S(\q,\w)$}
\begin{figure}[htb]
  \centering
  \includegraphics[width=0.9\linewidth]{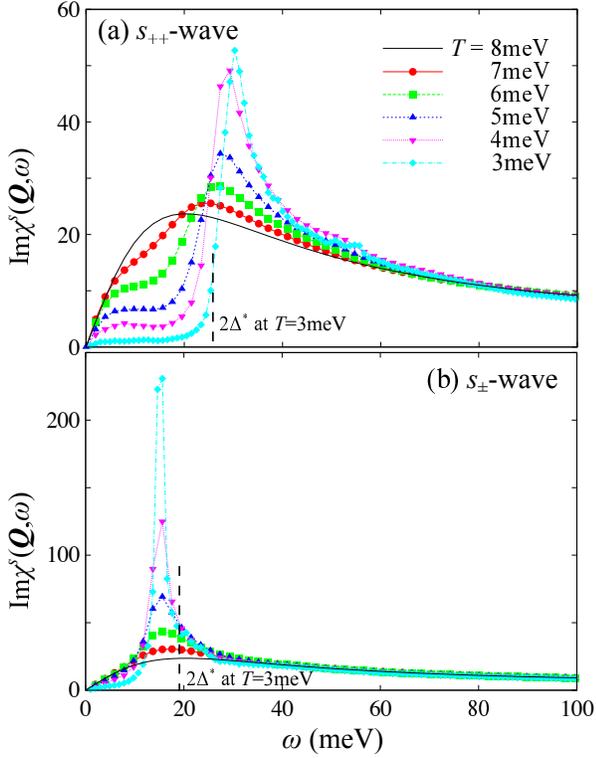}
  \caption{Energy dependence of ${\rm Im}\chi^S$ for 
(a)$s_{++}$ and (b)$s_{\pm}$ states below $T_{\rm c}$.
Black line($T_{\rm c}=8$meV) is the normal state.
In the $s_{++}$-wave state,
the height of the peak saturates at low temperatures.
}
  \label{resonance}
\end{figure}

Figure \ref{resonance} shows the $\w$-dependence of 
${\rm Im}\chi^S(\Q,\w)$ in the superconducting state
$T\le T_{\rm c}$ obtained by the FLEX approximation.
The lowest temperature is $T=3$meV, which is considerably 
lower than $T_{\rm c}=8$meV.
We set $\a_S(T_{\rm c})=0.95$.
In the case of $s_{++}$ state shown in Fig. \ref{resonance}(a),
the hump structure at $\w_{\rm res}\lesssim30$meV
becomes taller and sharper as $T$ is lowered.
The hump structure looks like a resonance peak at 
the lowest temperature $T = 3$meV.  
The energy position $\w_{\rm res}$ slightly increases as
$T$ decreases, and $\w_{\rm res}$ is slightly above $2\Delta^{\ast}$.
Figure \ref{resonance} (b) shows the result for $s_{\pm}$ state.
Compared to $s_{++}$ state, the magnitude of the structures 
are much larger and the resonance energy $\omega_{\rm res}$ does not move.

Compared to the result at $T=T_{\rm c}$, 
the height of ${\rm Im}\chi^S(\Q,\omega)$ in the $s_{++}$ state
is approximately twice in size,
while the height of the resonance peak in the $s_{\pm}$ state
is approximately 9 times as large.
In many Fe-based superconductors, the observed ``resonance peak'' 
is not so sharp, its weight is not so large,
and the height tends to saturate at low temperatures
\cite{keimer,neutron-NaFeAs}.
Thus, the obtained ${\rm Im}\chi^S(\Q,\omega)$ in the $s_{++}$ state 
well explains experimental results.

\begin{figure}[htb]
  \centering
  \includegraphics[width=0.99\linewidth]{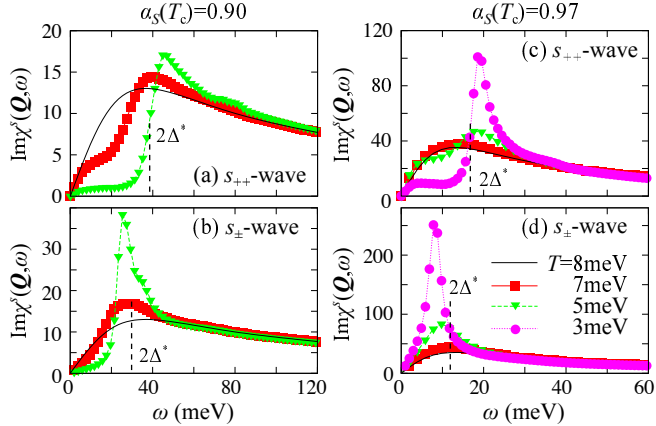}
  \caption{(a)(b)Energy dependence of ${\rm Im}\chi^S$ 
with $\alpha_S(T_{\rm c}) = 0.90$ for $T_{\rm c}\ge T \ge 5$meV: 
(a)$s_{++}$ and (b)$s_{\pm}$ states, 
respectively. 
(c)(d)Energy dependence of ${\rm Im}\chi^S$ with $\alpha_S(T_{\rm c})=0.97$
for $T_{\rm c}\ge T \ge 3$meV: 
(c)$s_{++}$ and (d)$s_{\pm}$ states, respectively.
In (c), the relation $\w_{\rm res} \approx 2\Delta^*$ holds even in 
$s_{++}$ wave state.
}
  \label{resonance9097}
\end{figure}

We also examine the spin susceptibility Im$\chi^S(\Q,\w)$
for $\alpha_S(T_{\rm c})=0.90$ and $0.97$
in Figs. \ref{resonance9097} (a)-(d).
When the system is away from the magnetic QCP ($\a_S(T_{\rm c})=0.90$),
the height of the hump structure in the $s_{++}$-wave state
becomes small as shown in Fig. \ref{resonance9097} (a).
On the other hand,
sharp resonance structure still exists in the $s_\pm$-wave state,
as shown in Fig. \ref{resonance9097} (b).

When the system is very close to the magnetic QCP 
($\a_S(T_{\rm c})=0.97$),
resonance-like peak structure is obtained in the $s_{++}$-wave state
in Fig. \ref{resonance9097} (c).
In the $s_\pm$-wave state,
resonance peak becomes very large 
shown in Fig. \ref{resonance9097} (d).
We note that peak structure in the $s_{++}$-wave state
for $\a_S(T_{\rm c})=0.97$ (\ref{resonance9097} (c))
is similar to that in the $s_\pm$-wave state for $\a_S(T_{\rm c})=0.90$ 
(\ref{resonance9097} (b)).
These results suggest that results from neutron-scattering experiments should be discussed carefully by considering the distance from the magnetic QCP.

\section{Analysis}

\subsection{Reason for the hump structure in $s_{++}$-wave state}
Here, we discuss the origin of the resonance or hump structure
in the dynamical spin susceptibility in the superconducting state.
Below, we drop the orbital degrees of freedom for simplicity.
The imaginary part of spin susceptibility $\chi^S$ can be written
in terms of real and imaginary parts of bare susceptibility.
We set $\Psi' = {\rm Re}(\chi^0 + \phi^0)$ and $\Psi'' = {\rm Im}(\chi^0 + \phi^0)$.
\begin{eqnarray}
  \chi^S &=& \frac{\Psi' + i\Psi''}{1-U(\Psi'+i\Psi'')} .
\end{eqnarray}
Taking the imaginary part of \(\chi^S\), we obtain
\begin{eqnarray}
  {\rm Im}\chi^S 
  &=& \frac{\Psi''}{(1-U\Psi')^2+(U\Psi'')^2} .
\end{eqnarray}
The denominator contains the term $U\Psi'$,
which at $\omega=0$ corresponds to the Stoner factor.

Here, we introduce the 
dynamical spin Stoner factor $\a_S(\w)$,
which is given by the largest real-part of the eigenvalue of 
$\widehat{\Gamma}^S\widehat{\Psi}(\q,\w)$
in the multiorbital model.
$\a_S(\w)$ at $\w=0$ is equivalent to $\a_S$
introduced in Sec. II.D.

\begin{figure}[htb]
  \centering
  \includegraphics[width=0.7\linewidth]{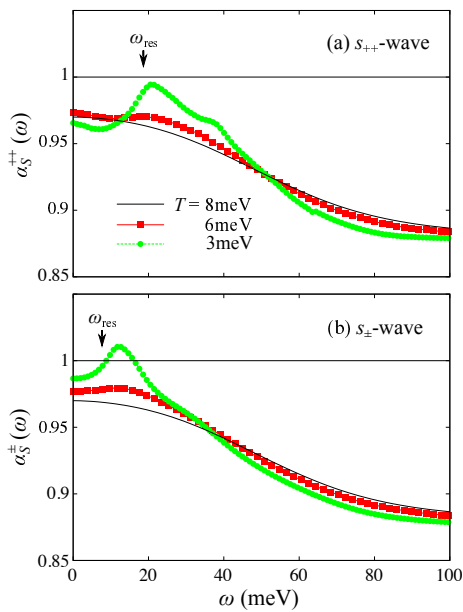}
  \caption{
    Dynamical spin Stoner factor $\a_S(\w)$ 
    given by the FLEX for (a)$s_{++}$ and (b)$s_{\pm}$ states.
    Here, $U$ is fixed under the condition $\a_{S}(T_{\rm c})=0.97$.
  }
  \label{acw-FLEX}
\end{figure}

Figure \ref{acw-FLEX} 
shows the dynamical spin Stoner factor 
for (a) $s_{++}$-wave state $\a_S^{++}(\w)$ and
that for (b) $s_\pm$-wave state $\a_S^\pm(\w)$,
given by the FLEX approximation.
In both cases, we set $\a_S=0.97$ at $T=T_{\rm c}$.
Both $\a_{S}^{++}(\w)$ and $\a_{S}^\pm(\w)$ 
take the maximum values at finite $\w$.
In the $s_\pm$-wave state, at $T=3$meV,
$\a_S^\pm(\w)$ reaches unity at $\w\approx12$ meV,
which corresponds to the resonance energy 
in the $s_\pm$-wave state shown in Fig. \ref{resonance} (b).
In the $s_{++}$-wave state, at $T=3$meV,
$\a_S^{++}(\w)$ reaches nearly unity at $\w\approx20$ meV,
which corresponds to the peak energy $\w_{\rm }$ 
in the $s_{++}$-wave state in Fig. \ref{resonance} (a).
Thus, the resonance-like peak structure
in the $s_{++}$-wave state originates from the 
condition $\a_S^{++}(\w)\approx1$ at $\w\approx\w_{\rm res}$.

\begin{figure}[htb]
  \centering
  \includegraphics[width=0.7\linewidth]{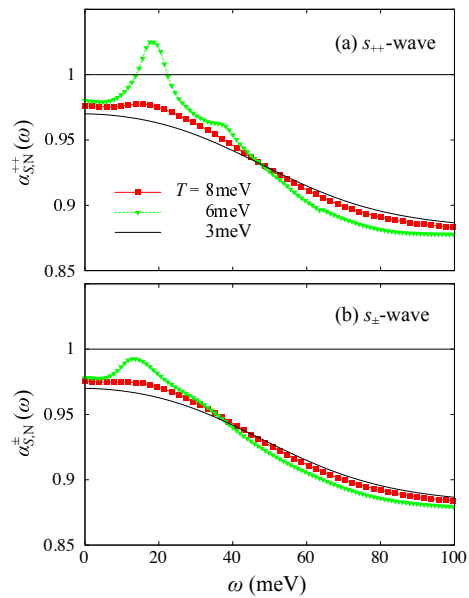}
  \caption{
    Normal-part dynamical Stoner factor 
    given by dropping ${\hat\phi}^0(q)$ in Eq. (\ref{phi0}),
    for (a)$s_{++}$ and (b)$s_{\pm}$ states.
    Here, $U$ is fixed under the condition $\a_{S}(T_{\rm c})=0.97$.
  }
  \label{acw-RPA}
\end{figure}

To understand the role of the coherence factor,
we also show the normal-part spin Stoner factors,
$\a_{S,{\rm N}}^{++}(\w)$ and $\a_{S,{\rm N}}^\pm(\w)$,
in Fig. \ref{acw-RPA}.
They are defined as
the largest real-part of the eigenvalue of
$\widehat{\Gamma}^S\widehat{\chi}^0(\q,\w)$ in the superconducting state.
(The coherent factor due to ${\hat\phi}^0(q)$ is dropped.)
We set $\a_{S}=0.97$ at $T=T_{\rm c}$.
Below $T_{\rm c}$,
both $\a_{S,{\rm N}}^{++}(\w)$ and $\a_{S,{\rm N}}^\pm(\w)$ 
increases with $\w$ for $\w\lesssim\w_{\rm res}$, 
reflecting the coherence peak
in the density-of-states at $\w=\pm\Delta^*$.
$\a_{S,{\rm N}}^\pm(\w)$ is smaller than $\a_{S,{\rm N}}^{++}(\w)$ 
at $\w\sim\w_{\rm res}$ since ${\rm Im}\Sigma(\k,\w)$
is larger in the $s_\pm$-wave state.
By making comparison between  Fig. \ref{acw-FLEX} and Fig. \ref{acw-RPA},
we find that the coherence factor enlarges (reduces)
the dynamical spin Stoner factor in the 
$s_\pm$-wave ($s_{++}$-wave) state.

We note that the authors in Refs.
\cite{mutou-hirashima,mutou-hirashima2}
studied the Kondo insulator model 
using the dynamical mean-field theory (DMFT),
and obtained large hump structure in $\chi^S(\q,\w)$
below the Kondo temperature.
The origin of the hump structure of $\chi^S$ in Kondo insulator,
which is actually observed in CeNiSn \cite{Kadowaki,Raymond},
is expected to be the same as that in the $s_{++}$-wave state, 
that is, the suppression of the inelastic scattering at low energies.

To summarize, we verified that $\a_S^\pm(\w)\approx1$ is realized 
in the $s_\pm$-wave state at the resonance energy $\w=\w_{\rm res}$.
The condition $\a_S^{++}(\w)\lesssim1$ at $\w=\w_{\rm res}$
is also realized in the $s_{++}$-wave state.
For $|\w|\gtrsim3\Delta^*$, $\a_S^\pm(\w)$ and $\a_S^{++}(\w)$
are suppressed by the large inelastic scattering $\gamma(\k,\w)$
\cite{onari-resonance,onari-resonance2}.
For this reason, Im$\chi^S(\Q,\w)$ shows resonance-like behavior 
even in $s_{++}$-wave state, when the normal state is close to 
the magnetic QCP.
We verified that, in the RPA without self-energy,
$\a_S^{++}(\omega)$ for $T<T_{\rm c}$ is smaller than 
$\a_S(\omega)$ at $T=T_{\rm c}$ for any $\omega$.
For this reason, the RPA fails to reproduce the hump
structure in the $s_{++}$-wave state
in Figs. \ref{comparison}-\ref{resonance9097}.

%
%

\section{Summary}

\begin{figure}[htb]
  \centering
  \includegraphics[width=0.9\linewidth]{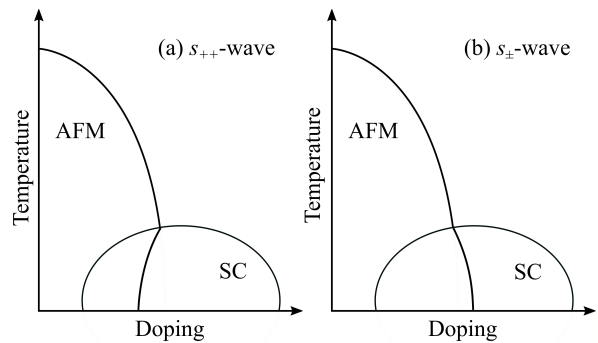}
  \caption{
    Schematic magnetic-superconducting phase diagram expected
    from the Stoner factor in the superconducting state shown
    in Figs .\ref{feedback} and \ref{feedback_comp}
    : (a) $s_{++}$-wave state, (b) $s_{\pm}$-waves state.
  }
  \label{phase-diagram}
\end{figure}

In this paper, we studied the dynamical spin susceptibilities
$\chi^S(\q,\w)$ in the $s_{++}$-wave and $s_\pm$-wave states
using the FLEX approximation.
We calculate the low-temperature electronic states ($T\ge1$meV)
accurately, by using very large number of 
Matsubara frequencies ($2^{16}$),
based on the multi-step FLEX method in Appendix A.
In this method, we reduce the memory size of
$\chi^0_{ll'mm'}(\q,i\w_l)$ and $\phi^0_{ll'mm'}(\q,i\w_l)$
by assigning crude $\k$-meshes for high Matsubara frequencies $\w_l$.
In the FLEX approximation in the superconducting state,
$\a_S$ is approximately independent of $T$ as shown in Fig. \ref{feedback}.
Near the magnetic QCP,
$\a_S$ slightly increases below $T_{\rm c}$ in $s_\pm$-wave state,
whereas it decreases for $T\ll T_{\rm c}$ in $s_{++}$-wave state.
This fact means that the expected phase diagrams
for  $s_{++}$-wave and $s_\pm$-wave states
do not have a pronounced difference,
as schematically shown in Figs. \ref{phase-diagram} (a) and (b).
This result is consistent with the phase diagram given by
the mean-field approximation in Ref. \cite{Tohyama-MF}

We also studied the energy dependence of 
${\rm Im}\chi^S(\q,\w)$ for 
$3{\rm meV} \le T \le 8{\rm meV} (=T_{\rm c})$
using the FLEX approximation.
Figure \ref{resonance} shows the numerical results 
in the case of $\a_S(T_{\rm c})=0.95$, 
which would correspond to the optimally-doped case.
Then, we obtain sharp peak structures in ${\rm Im}\chi^S(\q,\w)$ 
at $\w=\w_{\rm res}$ even in the $s_{++}$-wave state.
The relation $\w_{\rm res}\sim 2\Delta^*$ holds
in the $s_{++}$-wave state; see Fig. \ref{renorm}.
In the $s_\pm$-wave state,
resonance peak is very sharp and the resonance condition 
$\w_{\rm res}<2\Delta^*$ is satisfied.
Figures \ref{resonance9097} (a) and (b) show the 
results in the case of $\a_S(T_{\rm c})=0.90$,
which correspond to over-doped case.
Then, the peak structures in ${\rm Im}\chi^S(\q,\w)$ 
in the $s_{++}$-wave state becomes tiny.
In the $s_\pm$-wave state,
resonance peak is realized even in the over-doped case.
These results confirm the 
self-energy driven resonance-like peak in $s_{++}$-wave state,
which was proposed in our previous semi-microscopic study
\cite{onari-resonance,onari-resonance2}.
That is, the enhancement in $\chi^S(\q,\w)$ 
due to self-energy effect 
({\rm i.e.}, suppression of inelastic scattering below $T_{\rm c}$)
exceeds the suppression due to coherence factor effect.

Both the $\w$- and $T$-dependence of the
peak structure in the $s_{++}$-wave state near the magnetic QCP,
in Fig. \ref{resonance} (a) ($\a_S(T_{\rm c})=0.95$)
and Fig. \ref{resonance9097} (c) ($\a_S(T_{\rm c})=0.97$),
resemble to the experimental 
results in Na(Fe,Co)As and FeSe
reported in Refs. \cite{neutron-NaFeAs,neutron-FeSe}.
Thus, characteristic inelastic neutron spectra in 
optimally-doped Fe-based superconductors 
are well explained in the present FLEX study, 
if the $s_{++}$-wave superconducting state is assumed. 
As the system approaches to the magnetic QCP at $T=T_{\rm c}$,
which corresponds to the optimally-doped case,
Im$\chi^S(\q,\w)$ smoothly changes to resonance-like sharp peak structure
($\w_{\rm res}\approx 2\Delta^*$) in the $s_{++}$-wave state,
since $\a_S$ is very close to unity for $T\le T_{\rm c}$ 
in the FLEX approximation.

Since the overall ($\w/\Delta^*$)-dependence of 
${\rm Im}\chi^S(\Q,\omega)$ is insensitve to $\Delta^*$
for $T\lesssim 0.5T_{\rm c}$,
the present numerical results are reliable. 
We note that the resonance-like peak structure 
becomes sharper for smaller $\Delta^*$ at a fixed $\a_S$.

In the FLEX approximation, both the
coherence factor effect and self-energy effect 
on the dynamical susceptibility
are taken into account on the same footings.
In order to clarify the important role of the latter effect,
we perform the RPA analysis in Appendix B.
It is confirmed that the self-energy effect 
($(\w,T)$-dependence of the self-energy) discussed in Refs.
\cite{onari-resonance,onari-resonance2}
is indispensable 
for the resonance-like peak in the $s_{++}$-wave state
shown in Fig. \ref{resonance} (a) and 
Fig. \ref{resonance9097} (c).

\acknowledgements
We are grateful to S. Onari for useful discussions.
This work was supported by Grant-in-Aid for Scientific Research from 
the Ministry of Education, Culture, Sports, Science, and Technology, Japan.

\appendix
\section{Multi-step method of FLEX approximation}

In the present study, we have to perform the FLEX approximation 
at low temperatures ($T\ll T_{\rm c}$) accurately.
For this purpose, however, very large numbers of Matsubara frequencies
($-N_{\rm M}\sim N_{\rm M}$) are required in the numerical study.
To perform the FLEX at $T=1$meV precisely, for example,
$N_{\rm M}\sim2^{16}$ is required.
This fact has been preventing us from studying the FLEX approximation 
at very low temperatures.
To solve this difficulty, 
we introduced the multi-step FLEX method in the main text.
In this method, we reduce the memory size of
$\chi^0_{ll',mm'}(\q,i\w_l)$ and $\phi^0_{ll',mm'}(\q,i\w_l)$,
by assigning fine $\q$ meshes only for smaller $|\w_l|$.
We assign crude $\q$-meshes for larger $|\w_l|$
because the $\q$-dependence of $\chi^0(\q,i\w_l)$ is small then.

Here, we explain how to calculate the irreducible susceptibility 
$\chi^0(\q,i\w_l)$ based on the multi-step procedure.
First, we introduce the set of number of momentum meshes
and cutoff of Matsubara frequency number,
$\{(N_q^{(i)},N_{\rm M}^{(1)});i=1,2,\cdots\,L \}$.
Here, $N_q^{(i)}$ ($N_{\rm M}^{(i)}$) decreases (increases) with $i$.
For example, we may set $(N_q^{(1)},N_{\rm M}^{(1)})=(64^2,16)$,
$(N_q^{(2)},N_{\rm M}^{(2)})=(32^2,64)$, and
$(N_q^{(3)},N_{\rm M}^{(3)})=(16^2,256)$ for $L=3$.
Then, we introduce the following irreducible 
susceptibility from the $i$th energy width:
\begin{eqnarray}
\chi^{0(i)}(\q,i\w_l)&=&
-T\sum_n \sum_{\k}^{N_q^{(i)}}
G^{(i)}(\k+\q,i\e_n+\w_l)G^{(i)}(\k,i\e_n)
\nonumber \\
& &\times (\Theta_i(\e_n,\w_l)-\Theta_{i-1}(\e_n,\w_l))
\end{eqnarray}
where $G^{(i)}$ is the Green function
with meshes $(N_q^{(i)},N_{\rm M}^{(i)})$, 
$\Theta_i(\e_n,\w_l)=\theta(\e_{N_{\rm M}^{(i)}}-|\e_n|+\delta)
\theta(\e_{N_{\rm M}^{(i)}}-|\e_n+\w_l|+\delta)$, 
and $\Theta_{0} = 0$.
Then, the irreducible susceptibility in the 
multi-step RPA or FLEX is given as
\begin{eqnarray}
\chi^{0}=\sum_{i=1}\chi^{0(i)}
\end{eqnarray}
Note that $\chi^{0(i)}(\q,i\w_l)=0$ for $|\w_l| > \w_{N_{\rm M}^{(i)}}$.
(Since the $\q$-mesh number of $\chi^{0(i)}$ decreases with $i$,
interpolation should be performed for larger $i$.)
The obtained $\chi^{0}$ is very similar to that
given by the conventional RPA or FLEX
using $(N_q^{(1)},N_{\rm M}^{(L)})$, since
$\q$-dependence of $\chi^{0(i)}(\q,i\w_l)$ is small when $i$ or $|\w_l|$ is large.
In the same way,
we introduce the self-energy from the $i$th energy width
$\Sigma^{(i)}(\k,i\e_n)$, and then the total self-energy is given as
$\Sigma(\k,i\e_n)=\sum_{i=1}\Sigma^{(i)}(\k,i\e_n)$.

By employing this multi-step FLEX procedure, 
calculation time and memory can be saved.
In the present numerical research, 
we put $(N_q^{(1)},N_M^{(1)})=(128^2,16)$, 
$(N_q^{(L)},N_M^{(L)})=(2,2^{16})$, and $L=6$.
This multi-step procedure is justified from the 
basic idea of coarse graining or renormalization,
in that the momentum-dependence of physical quantities
are moderate for higher-energies.


\section{Dynamical spin susceptibility given by RPA}

In this paper, we studied the 
dynamical spin susceptibility $\chi^S$ in both $s_{++}$-wave
and $s_\pm$-wave states, by taking both the 
coherence factor effect and self-energy effect into account.
Both effects are comparably important 
in strongly-correlated superconductors
\cite{onari-resonance,onari-resonance2}.
For this purpose, we used the FLEX approximation,
in which both the self-energy and $\chi^S$
are calculated self-consistently.
Near the magnetic QCP,
resonance-like peak appears in $s_{++}$-wave state
since the enhancement in $\chi^S(\q,\w)$ 
due to self-energy effect 
({\rm i.e.}, suppression of inelastic scattering below $T_{\rm c}$)
exceeds the suppression due to coherence factor effect.

\begin{figure}[htb]
  \centering
  \includegraphics[width=0.8\linewidth]{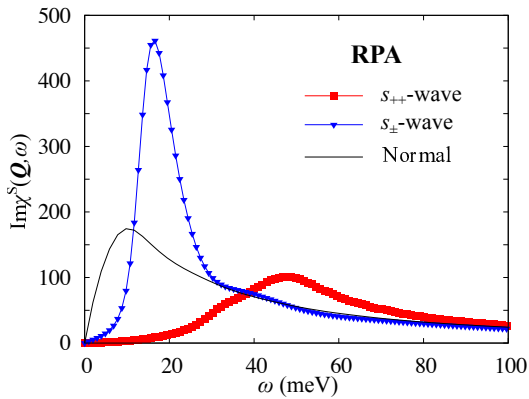}
  \caption{
Energy-dependence of Im$\chi^S$ given by the RPA
in the $s_{++}$-wave and $s_\pm$-wave states.
Im$\chi^S$ in the normal state is also shown.
  }
  \label{ApB1}
\end{figure}

Here, we perform the RPA analysis in order to explain that 
the resonance-like peak in the $s_{++}$-wave state
in Sec.III cannot be obtained once the 
self-energy effect is dropped.
Figure \ref{ApB1} shows the
energy-dependence of Im$\chi^S$ in both the 
superconducting states ($T=3\ {\rm meV}$)
and normal states ($T=8\ {\rm meV}$) given by the RPA.
We set $U=1.27$ eV for both normal and superconducting states,
$\Delta_0=20$ meV in the superconducting states.
The Stoner factor at $T=T_{\rm c}$ is 0.98.
In the RPA, 
the physical gap energy $\Delta_0^*$ 
discussed in the main text is equal to $\Delta_0$,
since the self-energy is absent.
In the $s_\pm$-wave state, clear resonance peak 
appear in Im$\chi^S$ at $\w\approx\frac12 (2\Delta_0)$,
consistently with previous RPA studies
\cite{maier-scalapino,eremin,Das-res,Kuroki-res,Korshunov-res,Korshunov-res2}.
In contrast, in the $s_{++}$-wave state, 
the obtained Im$\chi^S$ is smaller than that in the normal state
for $\w<2\Delta_0$.
In addition, the hump structure at $\w\gtrsim(2\Delta_0)$ is tiny.
Therefore, the resonance-like peak structure 
in the $s_{++}$-wave state obtained by the 
FLEX approximation, discussed in the main text,
cannot be obtained in the RPA.
Thus, the self-energy effect 
\cite{onari-resonance,onari-resonance2}
is indispensable 
for the resonance-like peak in the $s_{++}$-wave state.

\begin{figure}[htb]
  \centering
  \includegraphics[width=0.7\linewidth]{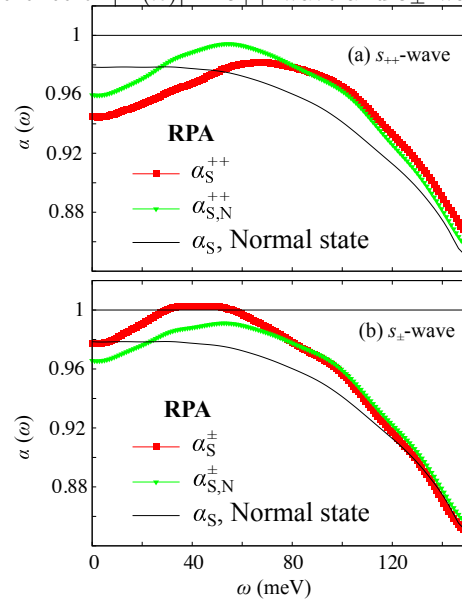}
  \caption{
Dynamical spin Stoner factor $\a_S(\w)$
and normal-part one $\a_{S,{\rm N}}(\w)$ given by the RPA,
for (a) $s_{++}$-wave and (b) $s_\pm$-wave states.
In $\a_{S,{\rm N}}(\w)$, the contribution from
${\hat\phi}^0(q)$ in Eq. (\ref{phi0}) is dropped.
  }
  \label{ApB2}
\end{figure}

In order to clarify the importance of the
coherence factor, we show both the 
dynamical spin Stoner factor $\a_S(\w)$ and
normal-part one $\a_{S,{\rm N}}(\w)$ given by the RPA
in Fig. \ref{ApB2}, 
for (a) $s_{++}$-wave and (b) $s_\pm$-wave states.
The difference between $\a_S(\w)$ and $\a_{S,{\rm N}}(\w)$
originates from the coherence factor given by 
${\hat\phi}^0(q)$ in Eq. (\ref{phi0}).
We see that $\a_{S,{\rm N}}(\w)$ is larger than 
$\a_{S}(\w)$ in the normal state at $\w\sim2\Delta_0$ meV,
by reflecting the coherence peak in the 
density-of-states at $\w=\pm\Delta_0$.
(In Fig. \ref{ApB2}, small difference between 
$\a_{S,{\rm N}}^{++}(\w)$ and $\a_{S,{\rm N}}^\pm(\w)$ 
originates from the difference of $|\Delta(\k)|$
in $s_{++}$-wave and $s_\pm$-wave states in Eq. (\ref{eqn:D0}).)
By including the coherence factor,
$\a_{S,{\rm N}}^{++}(\w)$ ($\a_{S,{\rm N}}^\pm(\w)$) 
becomes smaller (larger) than $\a_{S}(\w)$ in the normal state.

As we see in Fig. \ref{ApB2} (a),
the top of $\a_{S,{\rm N}}^{++}(\w)$ 
is comparable to that in the normal state.
Therefore, we cannot expect the emergence of 
the resonance-like peak structure 
in the $s_{++}$-wave state in the RPA.
In the FLEX approximation,
$\a_{S,{\rm N}}(\w)$ below $T_{\rm c}$ 
becomes much larger than that in the normal state,
as we show in Fig. \ref{acw-RPA}.
The reason is that the large $\gamma(\k,\w)$
in the normal state, which suppresses the spin susceptibility,
is reduced for $\w\lesssim3\Delta^*$ in the superconducting state.
This self-energy effect 
\cite{onari-resonance,onari-resonance2}
is indispensable 
for reproducing the resonance-like peak in
Im$\chi^S$ in the $s_{++}$-wave state.

\end{document}